\documentclass[a4paper,11pt]{article}
\usepackage{nfraconf,graphicx,cite}

\usepackage{times,pifont}

\begin{document}
\baselineskip=13pt

\title{The evolution of classical doubles: clues from complete samples}

\author{Katherine M.\ Blundell}
\address{Oxford University Astrophysics,
Keble Road, Oxford, OX1 3RH, UK\\
E-mail: kmb@astro.ox.ac.uk}

\author{Steve Rawlings}
\address{Oxford University Astrophysics,
Keble Road,
Oxford,
OX1 3RH, UK
}

\author{Chris J.\ Willott}
\address{Instituto de Astrofisica de Canarias,
38200 La Laguna, 
Tenerife,
Spain
}

\maketitle

\abstract{We describe the inter-dependence of four properties of
classical double radio sources --- spectral index, linear size,
luminosity and redshift --- from an extensive study based on
spectroscopically-identified complete samples.  We use these
relationships to discuss aspects of strategies for searching for
radio galaxies at extreme redshifts, in the context of possible
capabilities of the new generation of proposed radio telescopes.
}

\section{Introduction}
In the beginning, nearly\cite{Lai83}, was the 3C sample of radio
sources.  This was a very good sample of radio sources, except
that it suffered from a problem which afflicts each and every
single flux-limited sample: the luminosity--redshift ($P$--$z$)
degeneracy.  This tight correlation between luminosity and
redshift is illustrated schematically in Fig.\ {\ref{fig:pz}, for
the case of the 3C sample.

\begin{figure}[!h]
  \begin{center}
    \includegraphics[width=0.5\textwidth, angle = 90]{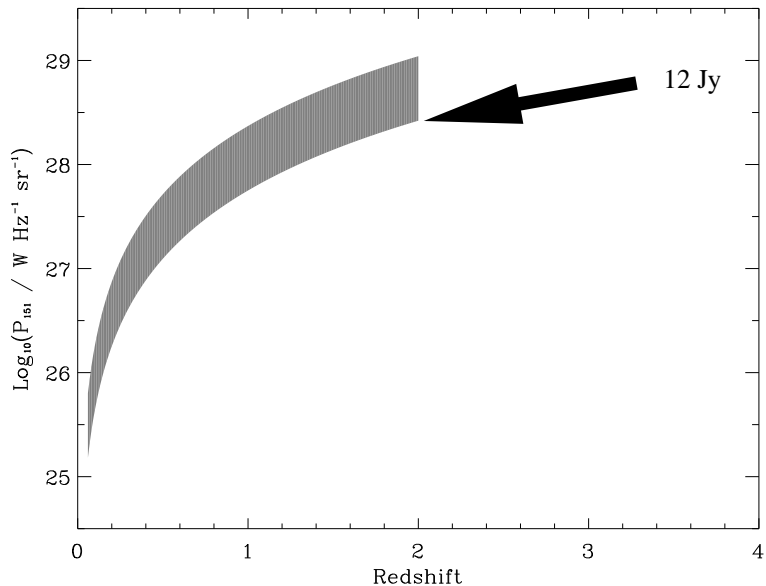}
    \caption{\small The shaded area represents the coverage of
    the $P$--$z$ plane by the 3C sample.  It demonstrates the
    tight correlation between luminosity and redshift inherent in
    any single flux-limited sample.  The lower boundary of the
    shaded region corresponds to a flux-limit of 12\,Jy at
    151\,MHz in a Universe with $\Omega_{\rm M} = 1$ and
    $\Omega_\Lambda = 0$, and assuming the radio sources have a
    low-frequency spectral index $\alpha = 1$, with $S_{\nu} 
    \propto \nu^{-\alpha}$, where $S_{\nu}$ is the flux density
    at frequency $\nu$.}
    \label{fig:pz}
  \end{center}
\end{figure}

In order to decouple radio source properties which primarily
depend on luminosity from those which depend on redshift, it is
necessary to increase substantially coverage of the $P$--$z$
plane over that occupied by the 3C sample.  We have done this by
pursuing a programme of identifying faint complete samples of
radio sources from 6C and 7C, selected at low frequencies similar
to 3C \cite{Raw98}.  By selecting samples of radio sources with
significantly lower flux-limits than 3C we find more
lower-luminosity objects than known before at high redshift, thus
breaking the $P$--$z$ degeneracy.  A {\sl combination} of
complete samples makes this possible.

\section{Linear-size evolution revisited}

It has been known for many years that the linear sizes of
classical double radio sources appear to become smaller with
increasing redshift \cite{Kap87}.  However, it has not been clear
until recently that this is primarily a dependence of linear size
on redshift rather than arising because of an anti-correlation of
linear size and luminosity \cite{Nee95,Blu99a}.

In fact the observed linear size evolution is {\sl an artifact of
the survey process itself}.  How strong this is found to be
depends crucially on the {\sl finding-frequency} of the samples
used in the study.  For example, when the finding-frequencies are
quite different for the brighter and the fainter samples (e.g.\
151\,MHz with a 12 Jy flux-limit and 408\,MHz with a 1\,Jy
flux-limit) as in the case of \cite{McC98}) then a very strong
evolution of linear size ($D$) with redshift is observed which
may be parameterised as $D \propto (1 + z)^{-3.5}$.  When the
finding-frequencies of the samples are the same, the linear size
evolution is found to be substantially milder \cite{Blu99a}, and
obeys the parameterisation $D \propto (1 + z)^{-1.3}$ .  The
mildness of the linear size evolution in this region of source
parameter space is very similar to that found \cite{Lac99} for
samples selected at 151\,MHz brighter than 12\,Jy and at 38\,MHz
brighter than 1.3\,Jy. 

The linear size a radio source has acquired when selected by a
survey depends on both its underlying jet-power and the
environmental profile into which it is expanding, and also on the
length of time elapsed since the jets first began expanding, that
is, on the {\sl age of the radio source}.

\section{What do hotspots do?}
\label{sec:palpha}
From the earliest \cite{Sch74,Bal82} to the most recent
\cite{Kai97,Blu99a} models of radio-source evolution, with
reasonable assumptions about the environments into which radio
sources expand, the luminosity of any individual radio source
decreases as time, or age, increases.  How dramatic this decrease
is does depend on the assumed density gradient \cite{Kai97} but
depends more dramatically on whether account is taken of the
r\^{o}le which hotspots play in a radio-source \cite{Blu99a}.
Hotspots do not reside in classical double radio sources merely
to facilitate measurement of angular sizes by the observer!
Rather, they play a key r\^{o}le in the evolution of a radio
source.  We have modelled two aspects of this r\^{o}le
\cite{Blu99a}.

First, the magnetic field of the hotspot causes significant
radiative losses on synchrotron particles during their dwell-time
in the hotspot, in the period of time before they are injected
into the lobe.  We were led to this by our observation that the
radio sources in our complete samples seemed to strongly indicate
that the more luminous a radio-source was, the steeper its {\sl
low-frequency spectrum} was (for example, when evaluated at
rest-frame 151\,MHz).  This rest-frame frequency regime is less
affected by synchrotron and inverse Compton losses than the
GHz regime and thus informs us on the energy distribution of
particles {\sl as initially injected} into the lobe of a
radio-source.  Modelling a steeper spectrum in a more powerful
source came naturally out of invoking stronger magnetic fields
(hence stronger radiative losses) in sources with higher
jet-powers.  By equating the jet-thrust and the pressure in the
compact hotspot, and invoking equipartition in the hotspot, we
find that the magnetic energy density in the hotspot is
proportional to the bulk kinetic power transported in the jet
(see equation 11 in \cite{Blu99a}).

Second, as a plasma element expands out of a compact hotspot into
a lobe whose pressure is considerably lower (and continues to
become lower as the radio source expands and gets older) it
suffers enhanced adiabatic expansion losses compared to the
expansion losses suffered by that element of plasma once it
continues to dwell in the lobe.  This is consistent with
observations of classical doubles which invariably show highly
compact hotspots --- embedded towards the outermost edges of the
smooth, low surface-brightness, extended emission which comprises
the lobe --- albeit with a bewildering menagerie of shapes and
structures \cite{Lea97,Har97,Bla92}.

\section{What governs the spectral indices at different
    frequencies?} 

In \S\ref{sec:palpha} we briefly described how a stronger hotspot
magnetic field will result in a steeper energy distribution of
particles being injected into the lobe, and hence a steeper
measured spectral index at low-frequency in the rest-frame.  This
model consistently explains the $P$--$\alpha$ correlation we
presented in \cite{Blu99a}.

Further losses cause a very different dependence of spectral
index when this is evaluated in the GHz regime.  In fact while we
found the low-frequency regime spectral index to correlate with
{\sl luminosity}, we found the GHz-regime spectral index to
correlate instead with {\sl redshift}.  This is illustrated in
Fig.\ \ref{fig:zalpha}.

\begin{figure}[!h]
  \begin{center}
    \includegraphics[width=0.7\textwidth]{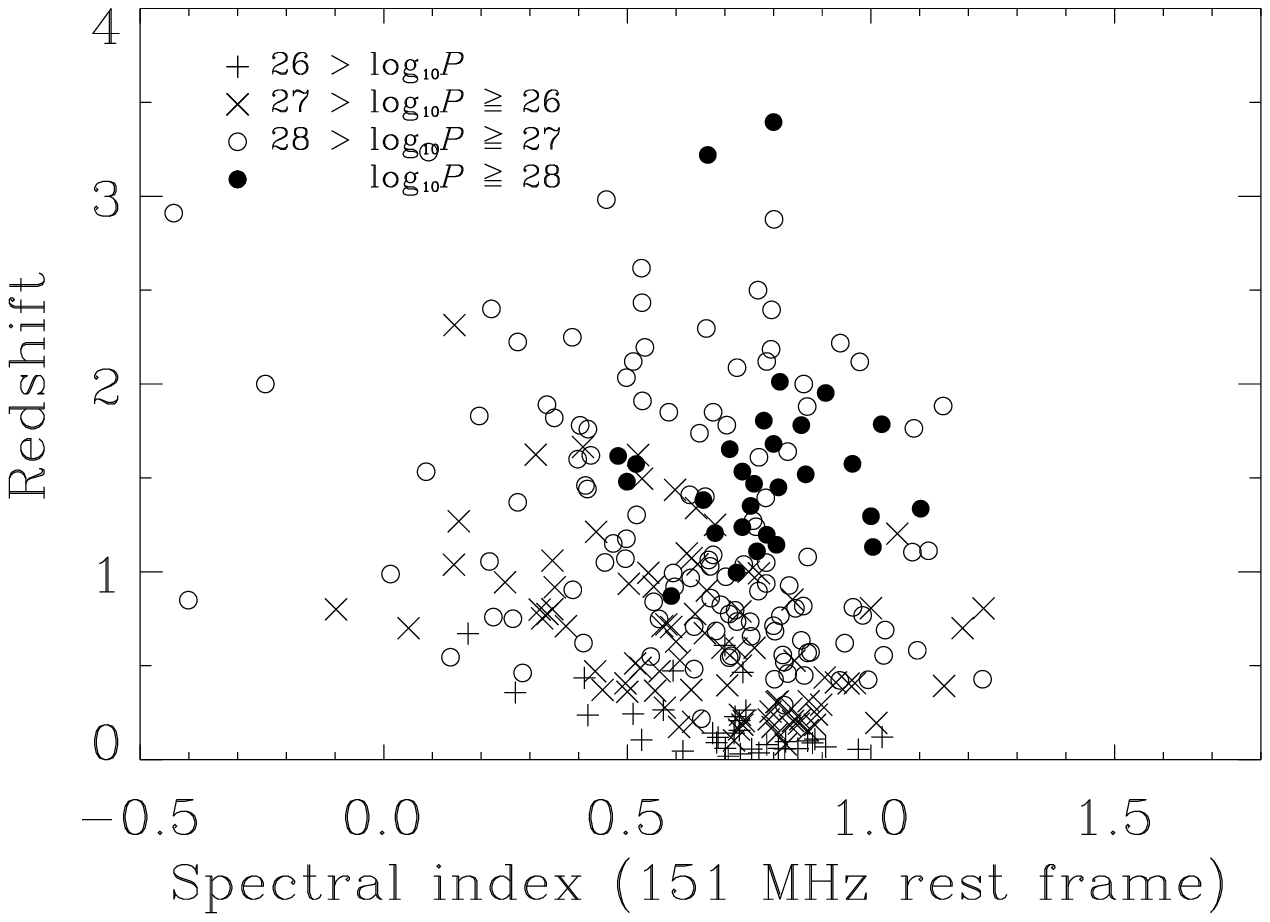}
    \includegraphics[width=0.7\textwidth]{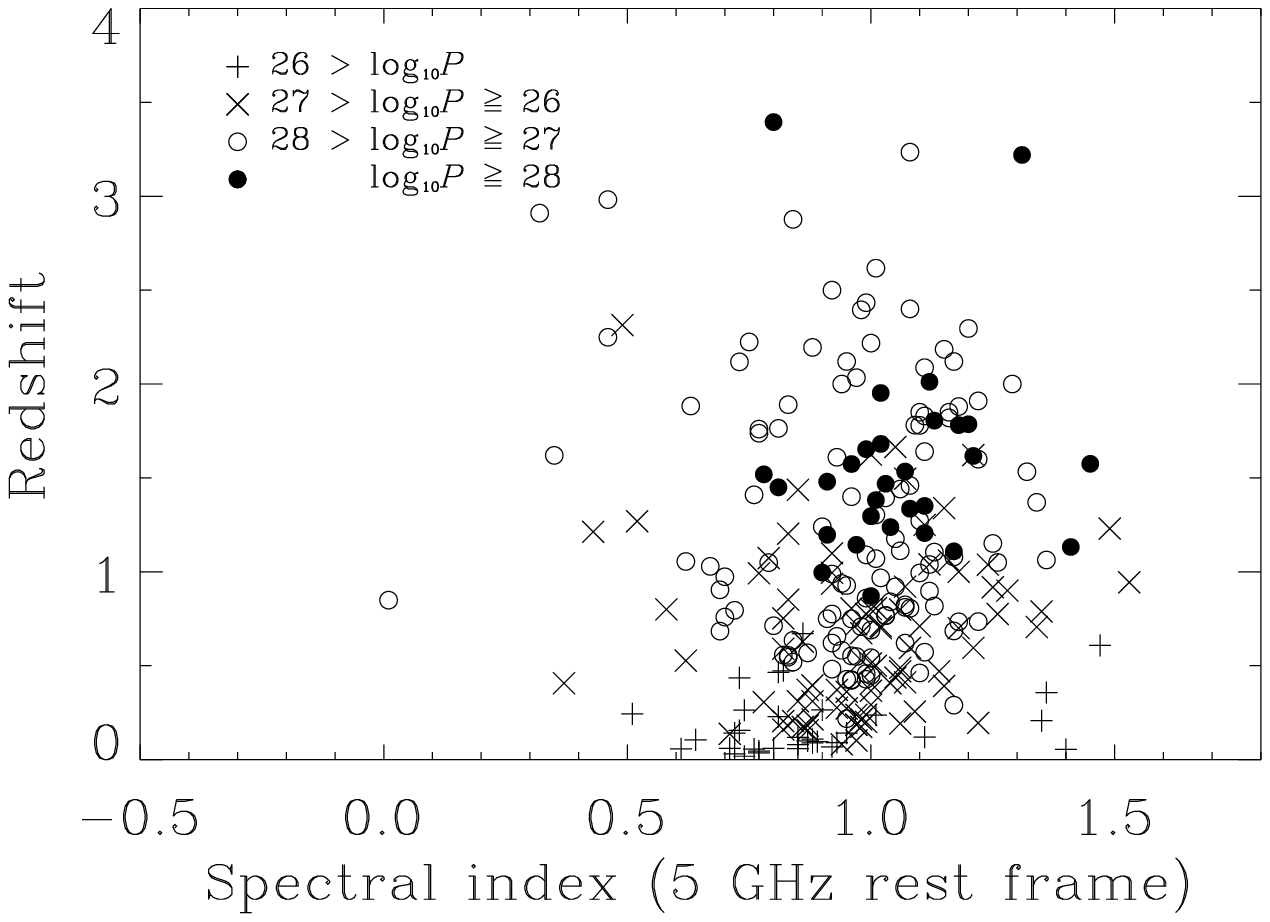}
    \caption{\small The upper panel plots the redshifts of the
    3C, 6C and 7C objects versus their spectral indices evaluated
    at 151\,MHz in their rest-frames.  There is no evidence for
    any independent correlation between these two properties.
    The lower panel shows that when the spectral index is
    evaluated at 5\,GHz in their rest-frames, this does
    significantly depend on redshift, indicating the increasing
    importance of inverse Compton losses in this frequency
    regime.  } \label{fig:zalpha} \end{center}
\end{figure}

The low-frequency spectral indices of lobes are also found to
increase with increasing linear size; this may be seen in Fig.\
\ref{fig:dalpha}.

\begin{figure}[!h]
  \begin{center}
    \includegraphics[width=0.7\textwidth,angle=90]{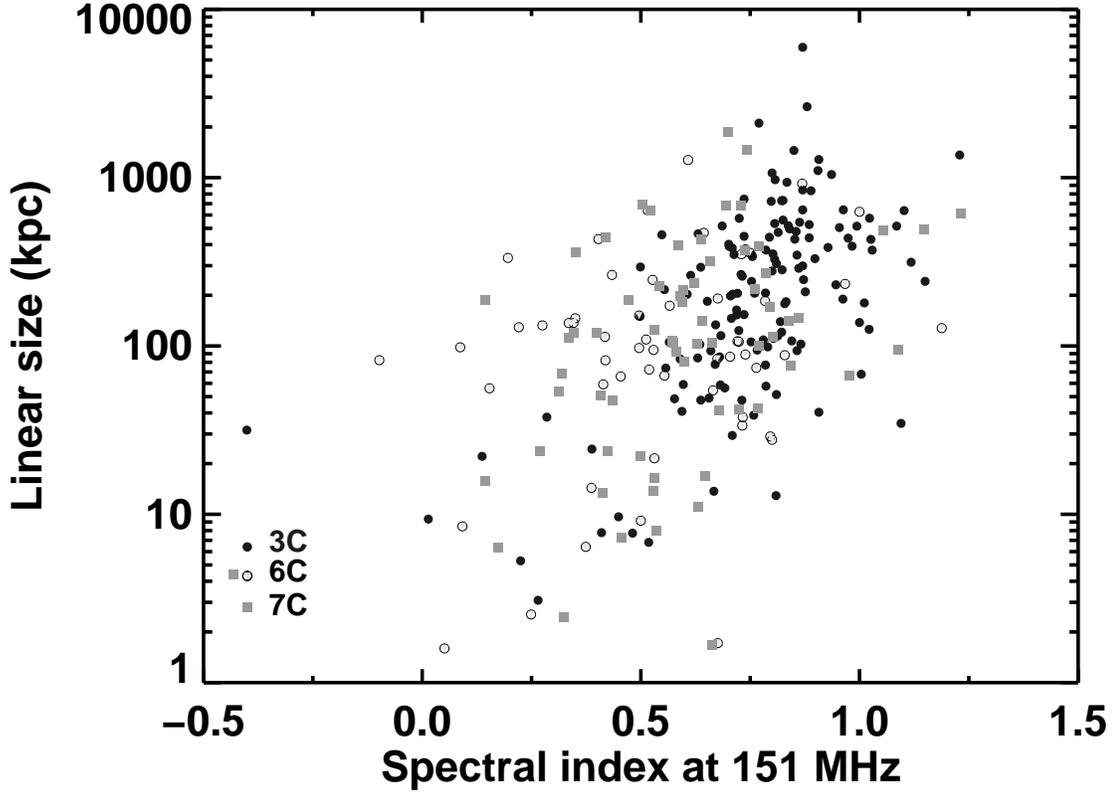}
    \caption{\small This plot shows the projected linear sizes of the
    3C, 6C and 7C objects versus their spectral indices evaluated
    at 151\,MHz in their rest-frames.  These quantities are seen to
    be significantly correlated.  Note that two competing effects
    are occuring: the $P$--$\alpha$ correlation, most manifest in
    the high-power sources and the $D$--$\alpha$ correlation,
    seen in the large linear-size, low-powered sources.      
    }
    \label{fig:dalpha}
  \end{center}
\end{figure}

We interpret the $D$--$\alpha$ correlation as being due to the
decrease in magnetic field as the lobes expand. For a fixed
observing frequency, the Lorentz factors of particles
contributing most of their emission at this frequency will be
{\sl higher} in a {\sl lower} magnetic field since the
frequency-dependence of synchrotron radiation from an electron
with Lorentz factor $\gamma$ in a magnetic field $B$ is given by:
\begin{equation}
\gamma = \biggl(\displaystyle\frac{m_{\rm e}}{eB} 2\pi \nu \biggr)^{
          \frac{1}{2}},
\end{equation}
where $m_{\rm e}$ is the rest mass of an electron and $e$ is the
charge on an electron.  

Thus as the radio-lobe expands and its magnetic field decreases
particles with a higher Lorentz factor are required to radiate at
the chosen observing frequency.  Given the power-law exponent of
typical energy distributions, the number of high Lorentz factor
particles is smaller than the number of less energetic particles.
Moreover, the adiabatic expansion of the lobes {\em per se},
while preserving the shape of the spectrum, will shift to lower
frequencies any features in the spectrum such as a break
frequency.

\section{Where are the large, powerful radio sources?}

Factoring in the r\^{o}le of the hotspot goes a long way to
solving a problem nearly three decades old (see e.g.\
\cite{Bal82}), namely the dearth of large {\sl and} powerful
classical doubles.  On the face of it this is puzzling:
intuitively more powerful radio-sources should expand faster than
those with lower jet-powers, so the laws of probability alone
suggest it should be easy to find large, powerful sources.  But
they are not there: sources with $P > 10^{27}\,{\rm
W\,Hz^{-1}\,sr^{-1}}$ and with $D > 1\,{\rm Mpc}$ are not found
in existing surveys (see Fig.\ \ref{fig:pd}).

\begin{figure}[!h]
  \begin{center} 
    \includegraphics[width=0.7\textwidth, angle=90]{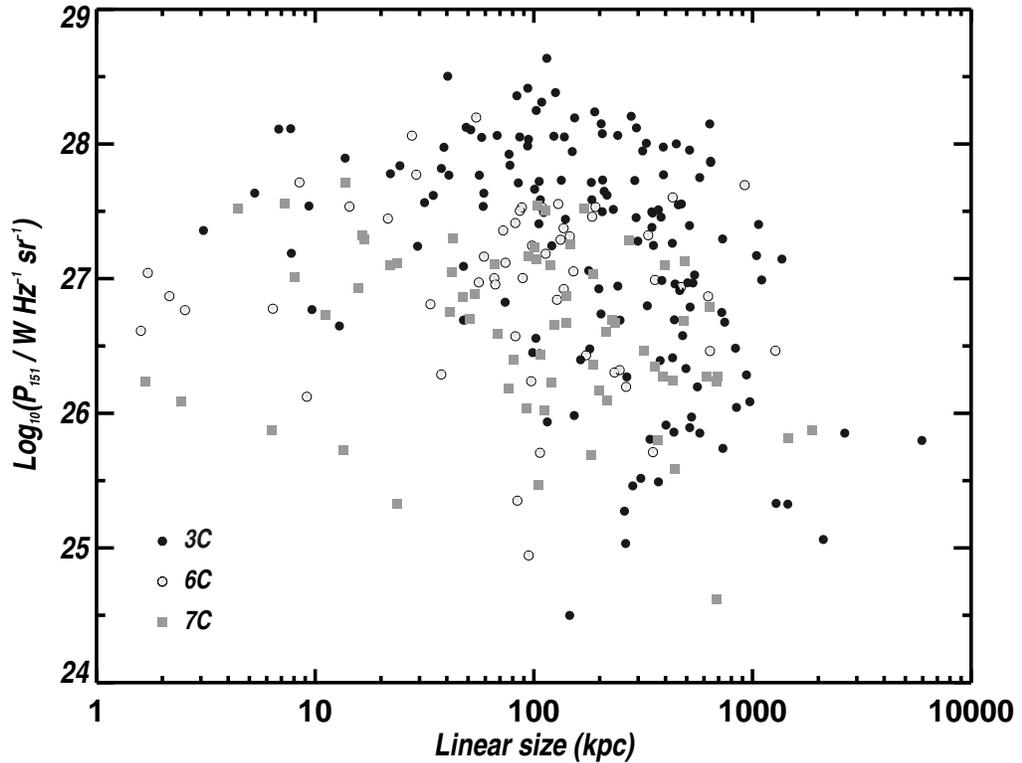} 
    \caption{\small The `$P$--$D$' plane for classical
    double radio sources from the 3C, 6C and 7C complete samples.
    Note the scarcity of sources with $P > 10^{27}\,{\rm
    W\,Hz^{-1}\,sr^{-1}}$ and with $D > 1\,{\rm Mpc}$.  }
    \label{fig:pd} \end{center}
\end{figure}

The explanation for their absence comes from consideration of how
the luminosity of a source with high jet-power evolves as it ages
given the influence of the hotspot, together with the consequence
of applying a flux-limit.  This is illustrated in Fig.\
\ref{fig:track}.  Note that inclusion of the influence of the
hotspot on the luminosity evolution of radio sources gives
steeply declining tracks and hence obviates the need to invoke
extremely steep density gradients in their environments, as
suggested by \cite{Bal82,Kai97}.

\begin{figure}[!h]
  \begin{center} \includegraphics[width=0.5\textwidth,
    angle=90]{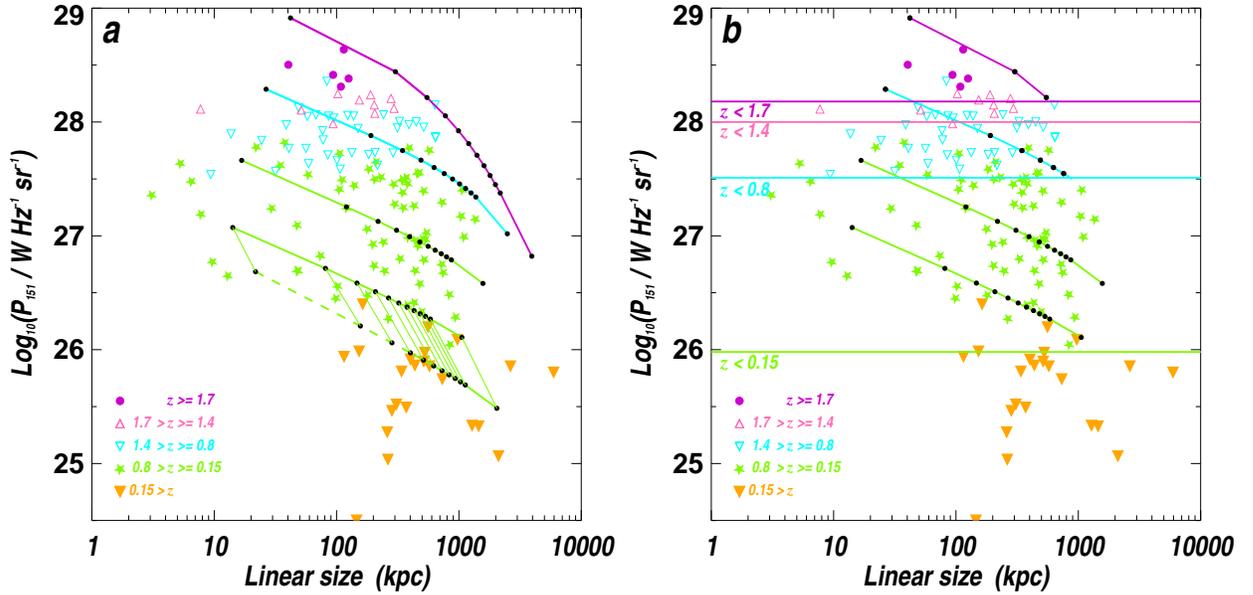} \caption{\small Overlaid on the
    `$P$--$D$' plane for the 3C sample in {\em \bf a} are model
    tracks tracing out the evolution of four example radio
    sources in luminosity and linear size, with from top to
    bottom $Q = 5 \times 10^{39}$ W at $z = 2$, $Q = 1 \times
    10^{39}$ W at $z = 0.8$, $Q = 2 \times 10^{38}$ W at $z =
    0.5$ and $Q = 5 \times 10^{37}$ W at $z = 0.15$.  The dashed
    line indicates how the lower track luminosity reduces by $<$
    half an order of magnitude if the ambient density becomes an
    order of magnitude lower.  In {\em \bf b} the horizontal
    lines represent the luminosities at which the flux-limit of
    12 Jy takes its effect at the different redshifts indicated.
    A combination of the dramatically declining
    luminosity-with-age of the high-$Q$ sources, their scarcity
    in the local Universe, together with the harsh reality of the
    flux-limit means that very powerful sources with large linear
    sizes are rarely seen.  } \label{fig:track} \end{center}
\end{figure}

\section{Young, distant radio galaxies}
\label{sec:yz}

The combination of flux-limits and declining luminosity evolution
point to an inevitable `youth-redshift degeneracy' in any
flux-limited sample.  In a recent letter to {\em Nature} we
examined the generality and ramifications of the youth-redshift
degeneracy (Blundell \& Rawlings 1999 \cite{Blu99b}).  We
described how one especial advantage of the youth-redshift
degeneracy is that very high-redshift sources, being inevitably
very young, are seen after only a short time ($< 10^7$\, yrs)
after their jet-triggering event.  A conclusion of that work was
that detection of high-redshift ($z > 4$) sources thus enables a
high-time-resolution study of triggering (and hence
galaxy-merging) rates within a Gyr of the Big Bang.

Given their unique importance as cosmological probes (compared
with say, the undateable optically selected high-$z$ quasars),
how then may we best find radio galaxies at extreme redshifts?

\section{Searches for distant radio galaxies}

In this conference we heard the exciting news of the discovery of
the first $z > 5$ radio galaxy (R\"{o}ttgering, these proceedings
and van~Breugel et al.\ 1999 \cite{van99}).  The fundamental
finding-frequency for the `filtered' sample from which this
object was found was 365\,MHz which of course samples emission at
2\,GHz in the rest-frame of this object.  Further filtering to
favour the detection of high-redshift objects was performed by
imposing a spectral index constraint on their sample of $\alpha >
1.3$ between 365\,MHz and 1.4\,GHz, and then further filtering
using the infra-red $K$-band relation with redshift
(\cite{Lil84,Eal96,Raw98}).  However, an object at $z = 5$ will
not be found by such a survey unless its GHz emission is
sufficiently luminous.  This requires that the source is
sufficiently young that inverse Compton and synchrotron losses
have not yet catastrophically depleted the emission in this GHz
regime.

A significant advantage in searching for high-$z$ radio galaxies
is achieved using a {\sl lower} fundamental
finding-frequency. This is because of the steep spectrum of these
objects and the avoidance of the frequency regime where inverse
Compton and synchrotron losses are most manifest.  Our own
filtered samples, 6C\ding{72} \cite{Blu98} (from which Rawlings
et al.\ found a $z = 4.4$ radio galaxy four years ago
\cite{Raw96}) and the on-going 6C\ding{72}\ding{72}, have as
their fundamental funding frequency 151\,MHz, the frequency at
which the 6C survey was undertaken by Hales et al.\ \cite{Hal88}.
For radio galaxies at $z = 5$ it is their emission sampled at
rest-frame 900\,MHz which determines whether they exceed the
flux-limit or not, and hence whether they will be members of the
survey.

Searches for high redshift radio galaxies invariably make use of
a spectral index constraint, since a steep spectral index is a
characteristic of known high-redshift radio galaxies.  This
arises partly because of the `$k$-correction' effect, partly
because the more distant radio sources are more luminous so the
$P$--$\alpha$ correlation takes effect and partly because of the
$z$--$\alpha_{\rm GHz}$ correlation which comes from increased
inverse Compton losses at high-redshift.

\begin{figure}[!h]
  \begin{center} 
	\includegraphics[width=\textwidth, bb= 33 123 589
	669]{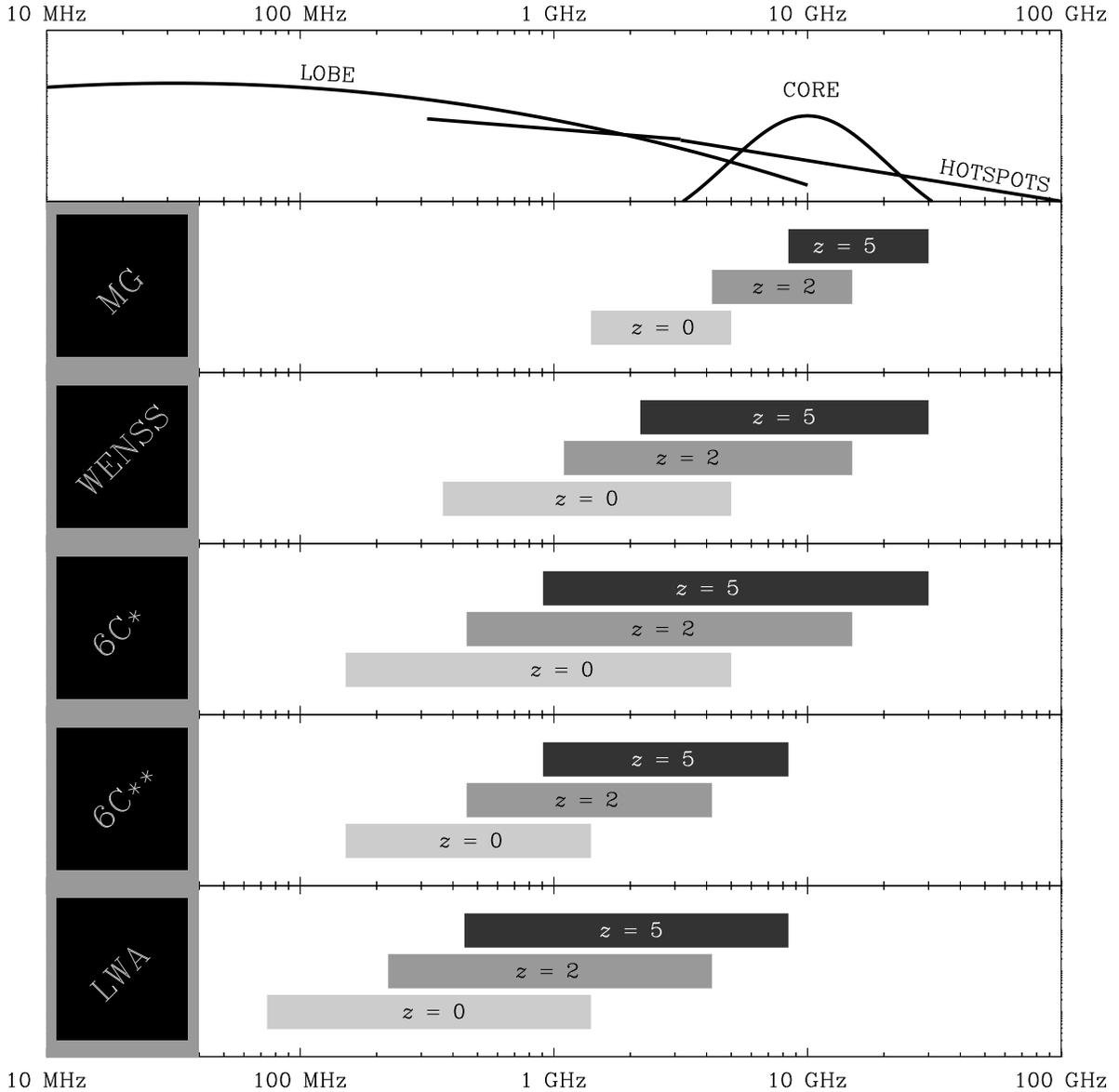}
        \caption{\small This figure indicates the different parts 
        of the rest-frame spectrum which are probed for objects
	at $z =	0$, $z = 2$ and $z = 5$, by different strategies
	for finding high-$z$ radio galaxies.  See text for more
	details.  } 
  \label{fig:hiz_hunt} 
  \end{center}
\end{figure}

Fig.\ \ref{fig:hiz_hunt} shows schematically the contribution of
the lobes, hotspots and core to the different frequency regimes
in the radio wave-band.  It also illustrates the different parts
of the rest-frame spectrum probed by four recent/on-going
searches for high-redshift galaxies.  The first is the MG survey
of Stern et al (1999), with the highest fundamental
finding-frequency reviewed here of 5\,GHz.  (Their spectral index
selection is made by cross-matching with a lower frequency of
1.4\,GHz.)  For $z = 5$ radio galaxies, their flux-limit at
5\,GHz probes emission at 30\,GHz in the rest-frame, where lobe
emission is highly depleted.  It is perhaps unsurprising then
that this sample, although it contains one previously known $z =
3.6$ radio galaxy, did not find any other $z > 3$ objects.

Surveys with a lower fundamental finding-frequency are much less
vulnerable to incompleteness: the WENSS survey of van~Breugel,
R\"{o}ttgering and collaborators which has found the first $z >
5$ radio galaxy, has its finding-frequency still in the GHz
rest-frame regime together with other filtering criteria
depending on the $K$--$z$ relation.  These make completeness
somewhat tricky to model.

Both the 6C\ding{72} and 6C\ding{72}\ding{72} samples have as
their fundamental finding-frequency 151\,MHz in the observed
frame.  Our selection criteria were chosen to minimize excessive
contamination from low-redshift interlopers {\sl without
excluding members of the target population}.  These precautions
are necessary for meaningful statements to be made about the
veracity of the purported `redshift-cutoff' in the co-moving
space density of radio galaxies (Jarvis et al.\ {\em in prep}).

The lowest plot indicates a survey which should be possible with
the currently evolving low frequency array concepts at NRL and
NFRA, LWA and LOFAR respectively. This plot indicates that part
of the spectrum which would be sampled by a finding-frequency of
75\,MHz (450 \,MHz in the rest-frame of a $z = 5$ radio galaxy).

A disadvantage of the 151-MHz 6C survey however, is its
relatively low spatial resolution (4\,arcmin).  The problem of
confusion is a significantly greater one in the deeper
6C\ding{72}\ding{72} sample than in the 6C\ding{72} sample.  This
then clearly identifies the two-pronged strategy which it is
hoped will shape the future capabilities of LWA/LOFAR.  The
unprecedented combination of {\sl low frequency \underline{and}
high-resolution is the key to the high-redshift Universe}.

The youth-redshift degeneracy, mentioned in \S\ref{sec:yz}, means
that there is a wide and increasingly ill-defined gulf between
the luminosity function (the co-moving space density of the
super-set of the radio-sources which make it above the various
survey flux-limits) and the birth function of radio sources.  We
have performed many Monte-Carlo runs of simulated complete and
filtered samples to constrain the birth function of radio
sources.  Comparison of these with our own observed complete and
filtered samples is the only means to constrain the
birth-function of radio sources throughout cosmic history
(Blundell, Rawlings \& Willott, in prep.).  LWA/LOFAR should
enable this to be constrained at the very earliest epochs that
classical double radio sources are triggered.

\end{document}